\documentclass[twocolumn,aps,prb,10pt,showpacs,raggedbottom,nobalancelastpage,amssymb,superscriptaddress,floatfix]{revtex4-1}
\usepackage{amsmath,amssymb}
\usepackage{txfonts}
\usepackage{hyperref}
\usepackage{braket}
\usepackage{graphicx}
\usepackage{hyperref}

\hypersetup{colorlinks=true, linkcolor=blue, citecolor=blue, urlcolor=blue}

\renewcommand{\Re}{\operatorname{Re}}

\newcommand\numb{\addtocounter{equation}{1}\tag{\theequation}}
\newcommand{\cH}{\mathcal{H}}
\newcommand{\cU}{\mathcal{U}}

\begin{document}

\title{Charge pumping in strongly-coupled molecular quantum dots}
\author{Patrick Haughian}
\affiliation{Physics and Materials Science Research Unit, University of Luxembourg, 1511 Luxembourg, Luxembourg}
\author{Han Hoe Yap}
\affiliation{NUS Graduate School for Integrative Sciences and Engineering, Singapore 117456, Republic of Singapore
}
\author{Jiangbin Gong}
\affiliation{NUS Graduate School for Integrative Sciences and Engineering, Singapore 117456, Republic of Singapore
}
\affiliation{Department of Physics, National University of Singapore, Singapore 117551, Republic of Singapore}
\author{Thomas L. Schmidt}
\email{thomas.schmidt@uni.lu}
\affiliation{Physics and Materials Science Research Unit, University of Luxembourg, 1511 Luxembourg, Luxembourg}

\date{\today}

\begin{abstract}
The interaction between electrons and the vibrational degrees of freedom of a molecular quantum dot can lead to an exponential suppression of the conductance, an effect which is commonly termed Franck-Condon blockade. Here, we investigate this effect in a quantum dot driven by time-periodic gate voltages and tunneling amplitudes using nonequilibrium Green's functions and a Floquet expansion. Building on previous results showing that driving can lift the Franck-Condon blockade, we investigate driving protocols which can be used to pump charge across the quantum dot. In particular, we show that due to the strongly coupled nature of the system, the pump current at resonance is an exponential function of the drive strength.
\end{abstract}

\maketitle

\section{Introduction}
Recent progress in fabrication and measurement techniques has broken new ground in the field of nanoscale physics: Currently realizable systems allow for the experimental examination and manipulation of single quantum states, putting within reach a wide variety of novel effects.\cite{vdz12,tao06} Specifically, the transport properties of such systems are the subject of extensive study, on the quest for pioneering designs of electronic components and circuitry.

One of the core promises of nanoelectronics is to accurately generate and control small amounts of current. In such a context, the paradigm of quantum pumping\cite{pot92,swi99,mos11} has received considerable attention. Charge pumping signifies a nonzero time-averaged flow of current through a quantum system as a result of the temporal variation of one\cite{san11,tor11} or several\cite{fio08,pra09,arr08} system parameters, even in the absence of voltage bias. This effect can be achieved, for instance, in a system between two leads at equal chemical potential, where the system parameters are modulated by an external AC signal. Nanoscale charge pumps have potential uses as sources of quantized, tunable currents.\cite{wri09,ste15}

The current generated by a quantum pump generally depends in a significant fashion on the drive protocol, which can manifest in various ways: For example, if the drive period is longer than the time scales inherent in the system, an adiabatic time evolution of the system can be used to obtain rather general results for the current.\cite{mos02,zho03,mos04} In contrast, for comparatively fast driving, the situation is less straightforward and the resulting pump current tends to depend strongly on the excited state spectrum of the system in question, as well as on the specific driving protocol that is being employed.\cite{str05,kae15}

Charge pumping has been studied frequently in electronic systems. However, nanoscale physics is not limited to electronics alone. In particular, the interactions between charges and optical or mechanical degrees of freedom open up further avenues for exploration.\cite{xia13,fre12,wal04} A prototypical nanoelectromechanical system (NEMS) exists for instance in the form of carbon nanotubes (CNT).\cite{sap06,lai15,ila10,hue09,fle06,las09} It has recently become possible to use electronic gates to define a quantum dot on a CNT and to tailor the interaction of electrons and quantized vibrational modes (``vibrons'') of the nanotube.\cite{wai13,ben14,let09}

In particular, the interplay between electronic and mechanical degrees of freedom in a NEMS can have profound consequences for its conductive properties: The simplest such system -- a single vibrational mode interacting with an electronic level -- already gives rise to an infinite ladder of composite electromechanical states (``\mbox{polarons}'') that can in principle contribute to conduction. In the limit of strong electron-vibron coupling, the transitions involving low-lying states of this ladder are exponentially suppressed, leading to a drastic reduction in current, a phenomenon called Franck-Condon blockade (FCB).\cite{koc04,koc05,koc06} On the other hand, it was shown recently that an AC gate voltage in resonance with the vibration can be used to actuate conduction channels that are much less strongly suppressed, which lifts the Franck-Condon blockade exponentially in the drive amplitude.\cite{hau16} In Ref.~[\onlinecite{hau16}], it was proposed to observe this effect in a CNT quantum dot, since such a system exhibits the required strong coupling, and the AC gate voltage could be supplied by the gates used to define the quantum dot.\cite{saz04} Here, we extend this setup by using additional gates to also modulate the coupling of the quantum dot to the lead electrodes. The availability of more than one time-dependent parameters then allows us to build a bridge towards charge pumping.

In this article, we study the current response of a Franck-Condon-blockaded quantum dots to several periodic drives. We consider a model for a quantum dot with coupled vibrational and electronic sectors, weakly coupled to a pair of metallic leads. A drive protocol is defined which modulates both the coupling to the leads and the energy level of the dot. As a result of this drive, we find that charge pumping through the dot can be achieved. Interestingly, we find that for drive frequencies resonant with the vibron mode, the pump current depends exponentially on the drive amplitude.

The paper is organized as follows: In Sec.~\ref{sec_model}, we lay out the model used to describe a doubly-driven electromechanical quantum dot. Sec.~\ref{sec_greens} contains the derivation of the current through the system, where we use the Keldysh nonequilibrium Green's function method in conjunction with the polaron tunneling approximation and a Floquet expansion. We apply this method to a specific driving protocol in Sec.~\ref{sec_pumping}, leading to a description of charge pumping. In Sec.~\ref{sec_conclusion}, we summarize our findings and discuss extensions and ideas for application.

\section{Model}
\label{sec_model}
\begin{figure}[t]
     \includegraphics[width=0.99\columnwidth]{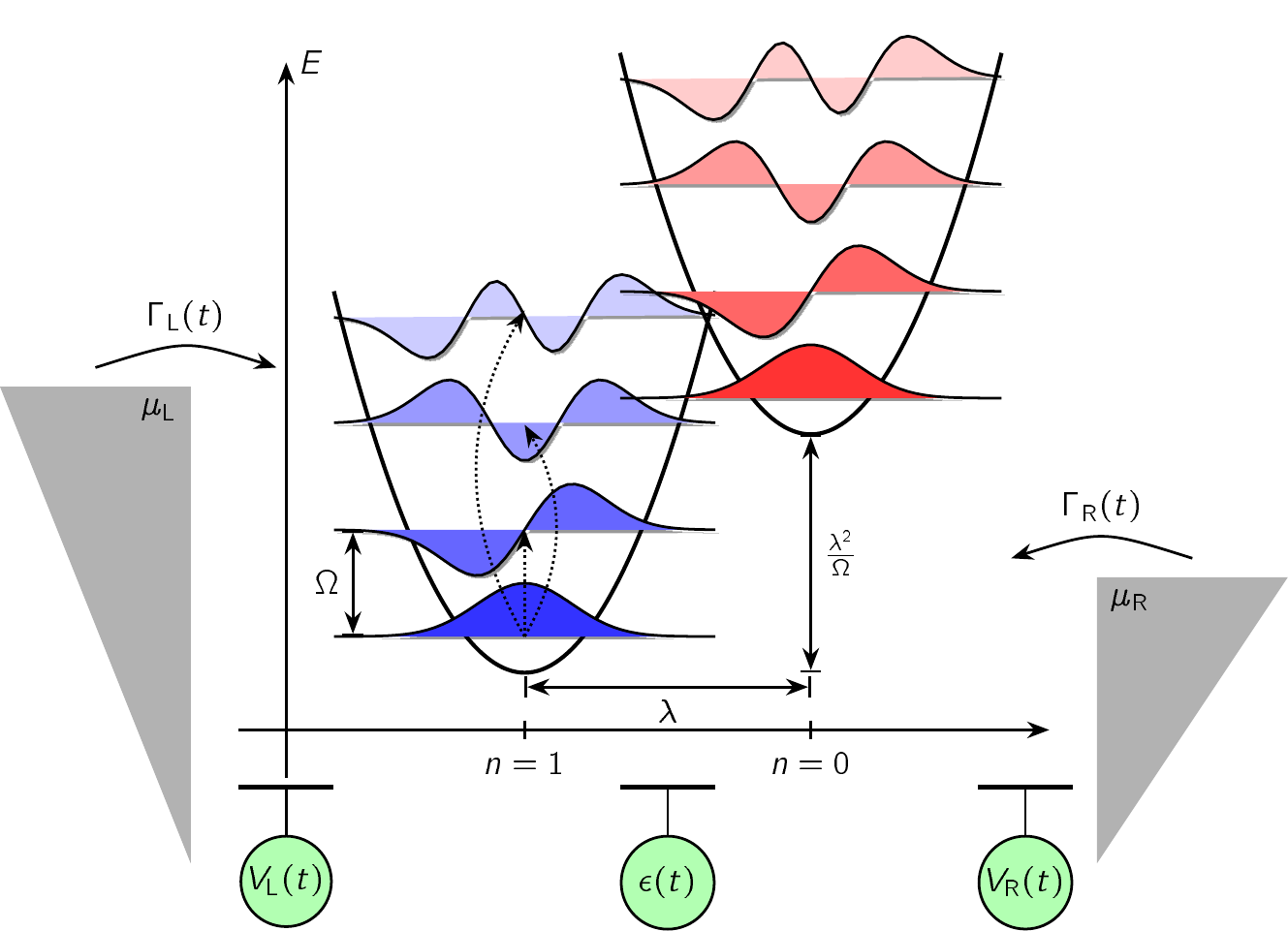}
 \caption{Schematics of the Hamiltonian from Eq.~\eqref{hamiltonian}. As a result of electron-vibron coupling, the vibrational spectrum has different ground state energies depending on the electron number \(n=\big\langle d^\dagger d\big\rangle\), with the unoccupied sector (\(n=0\)) having an energy increase of \(\lambda^2/\Omega\). The energy of the dot electron \(\epsilon(t)\) and the couplings to the leads \(V_\text{L, R}(t)\) are subject to periodic driving. }
\label{FCB}
\end{figure}

In principle, the electronic interactions on a quantum dot can be very complex as different vibron modes may couple to the charges in different electronic orbitals. To capture the essential physics, we use as a minimal model for electron-vibron interactions in the following the Anderson-Holstein Hamiltonian \({\cH=\cH_\text{dot}+\cH_{\text{lead}}+\cH_{\text{tun}}}\), where
\begin{align*}
\label{hamiltonian}
\cH_\text{dot}&=\Omega a^{\dagger}a+\bar{\epsilon}(t) d^{\dagger}d+\lambda(a^{\dagger}+a)d^{\dagger}d,\\
\cH_\text{lead}&=\sum_{\alpha=\text{L}, \text{R}} \sum_{k} \epsilon^{\phantom\dagger}_{k\alpha}c^{\dagger}_{k\alpha}c^{\phantom\dagger}_{k\alpha},\\
\cH_\text{tun}&=\sum_{\alpha=\text{L}, \text{R}} \sum_{k}\left[V_{k\alpha}(t)d^\dag c_{k\alpha}+\text{h.c.}\right],\numb
\end{align*}
denote the dot, lead and tunneling Hamiltonians, respectively. The quantum dot is modeled by a single electron level at energy \(\bar{\epsilon}(t)\), subject to a time-dependent gate voltage, and represented by the creation and annihilation operators \(d\) and \(d^\dag\) obeying the anticommutation relation \(\{d,d^\dag\}=1\). The vibrational sector of the dot is given by a single vibron mode of frequency \(\Omega\), described by bosonic operators \(a\) and \(a^\dag\). We set \(\hbar=1\) throughout. Electrons and vibrons interact via a coupling of the vibron displacement operator to the electron number, where \(\lambda\) denotes the strength of this coupling. The left and right leads are modeled as reservoirs of free fermions with energies \(\epsilon_{k\alpha}\), with the corresponding operators \(c_{k\alpha}\) and \(c^\dag_{k\alpha}\) obeying \(\{c_{k\alpha},c^\dag_{k'\alpha'}\}=\delta_{kk'}\delta_{\alpha\alpha'}\). Finally, the dot couples to each mode in the leads via the time-dependent tunneling amplitude \(V_{k\alpha}(t)\). A summary of the components of the Hamiltonian is given in Fig.~\ref{FCB}.

The dot Hamiltonian can be mapped onto a non-interacting model \cite{lan63} by applying the  unitary transformation \({\tilde{\cH}=\cU\cH\cU^{-1}}\), where \({\mathcal{U}=\exp[\lambda(a^{\dagger}-a)d^{\dagger}d/{\Omega}]}\). Indeed, this procedure results in a quadratic Hamiltonian \({\tilde{\cH}_{0}=\tilde{\cH}_{\text{dot}}+\cH_{\text{lead}}}\), with
\(\tilde{\cH}_\text{dot}= \epsilon(t)d^\dag d +\Omega a^{\dagger}a\), where the dot electron energy has been renormalized to \(\epsilon(t)=\bar{\epsilon}(t)-\lambda^2/\Omega\). While the lead Hamiltonian remains unchanged, the tunneling amplitudes are dressed by an exponential vibrational factor,
\begin{align}
\tilde{\cH}_\text{tun}=\sum_{\alpha} \sum_k \left[V_{k\alpha}(t)X^\dagger d^\dag c_{k\alpha}+\text{h.c.}\right],
\end{align}
where \(X=\text{e}^{-\frac{\lambda^2}{\Omega^2}\left(a^\dagger-a\right)}\) encodes the modification of the tunneling process due to the polaron.

In the limit of slow tunneling, \(V_{k\alpha}\ll\min{(\lambda,\Omega)}\), insight into the transport properties of the system has been obtained: In the time-independent case, \(V_{k\alpha}(t)\equiv V_{k\alpha}\), \(\epsilon(t)\equiv\epsilon\), the electron-vibron interaction leads to an exponential suppression of DC current, \(\langle I\rangle\propto\text{e}^{-\lambda^2/\Omega^2}\). This phenomenon is known as Franck-Condon blockade \cite{koc04,koc05,koc06} and can be pictured as follows: As a consequence of electron-vibron interaction, the lattice structure of the dot will be deformed in the presence of an electron, whereby a polaron is formed. If current is to flow through the dot, e.g., if the electron is to tunnel out, this composite state has to be broken up, which is energetically costly for strong electron-vibron interaction. Application of an AC voltage to the dot energy can supply the energy required to break up the polaron, thus facilitating electron tunneling and lifting the current blockade.\cite{hau16} In the following, we set up a formalism that allows us to treat periodic, resonant drives in both the dot energy \(\epsilon(t)\) and the coupling to the leads \(V_{k\alpha}(t)\).

\section{Floquet Green's functions}
\label{sec_greens}

\subsection{Polaron tunneling approximation}
\begin{figure}[t]
     \includegraphics[width=0.99\columnwidth]{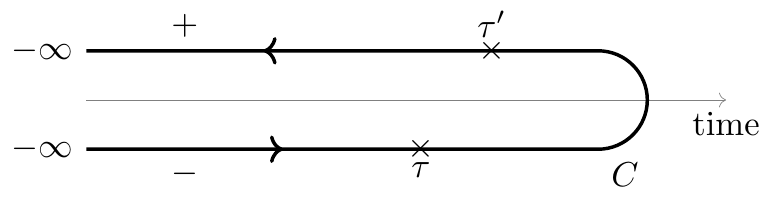}
 \caption{Keldysh integration contour \(C\), running from \(\tau=-\infty\) past the larger of the times \(\tau\) and \(\tau'\) in the lower half-plane, before returning to \(-\infty\) in the upper half plane.}
\label{contour}
\end{figure}

In order to calculate the current through the quantum dot, we make use of the nonequilibrium Green's function technique. The operator describing the charge current flowing through the lead \(\alpha\) is given by
\begin{align*}I_{\alpha}&=e\frac{\text{d}}{\text{d}t}\sum_{k}c^\dag_{k\alpha}c^{\phantom\dag}_{k\alpha}=-\text{i}e\left[\sum_{k}c^\dag_{k\alpha}c^{\phantom\dag}_{k\alpha},\tilde{\cH}\right]=\\
&=\text{i}e\sum_{k}V_{k\alpha}(t)X^\dag(t) d^\dag(t) c_{k\alpha}(t)+\text{h. c.},\numb
\end{align*}
where the time dependence of the operators is understood to arise from evolution with the full Hamiltonian \(\tilde{\cH}\). Therefore, the current expectation value can be expressed as
\begin{align}
\label{current_ev}
\left\langle I_\alpha(t)\right\rangle=e\sum_{k}V_{k\alpha}(t)F^{-+}(t,t)+\text{c.c.},
\end{align}
with \(F(\tau,\tau')=-\text{i}\big\langle T_{C} c_{k\alpha}(\tau')X^\dag(\tau') d^\dag(\tau') \big\rangle\) denoting the contour ordered ``mixed'' dot-lead Keldysh Green's function. Specifically, the times \(\tau\) and \(\tau'\) lie on the Keldysh contour \(C\), as depicted in Fig.~\ref{contour}. Transforming to real times \(t\) and \(t'\), one obtains the Keldysh \(2\times2\) matrix structure, defined by whether the times \(\tau\) and \(\tau'\) lie on the upper or lower half plane, respectively.

In principle, the Green's function \(F(\tau,\tau')\) can be calculated perturbatively in \(\tilde{\cH}_{\text{tun}}\) in the limit of weak coupling between dot and lead.
Unfortunately, a direct resummation of the perturbation series is impossible because \(\tilde{\cH}_{\text{tun}}\) contains the vibrational operator \(X\), and thus gives rise to expectation values for which Wick's theorem does not hold. Therefore, we use the polaron-tunneling approximation introduced in Refs.~[\onlinecite{mai11}]~and~[\onlinecite{hau16}], noting that in the regime of strong electron-vibron interaction the lifetime of the polaron, an estimate for which is given by the inverse of the energy shift \(\bar{\epsilon}-\epsilon=\lambda^2/\Omega\), will be much larger than the average tunneling time between dot and lead, which scales as \(|V|^2\). Hence it can be assumed that the vibrational sector of the dot will relax between each pair of tunneling processes. Within this approximation, the renormalized polaron dot Green's function \({D(\tau,\tau')=-\text{i}\big\langle T_{C}d(\tau)d^\dag(\tau')X(\tau)X^\dag(\tau')\big\rangle}\) obeys the Dyson equation
\begin{align*}
\label{dyson_time}
D(\tau,\tau')&=D_0(\tau,\tau')\\
&+\int_C\text{d}\sigma_1\int_C\text{d}\sigma_2 D_0(\tau,\sigma_1)\Sigma(\sigma_1,\sigma_2)D(\sigma_2,\tau'),\numb
\end{align*}
where the bare polaron dot Green's function reads \begin{align}
D_0(\tau,\tau')=-\text{i}\left\langle T_{C}d(\tau)d^\dag(\tau')\right\rangle_0 \left\langle T_C X(\tau)X^\dag(\tau')\right\rangle_0\end{align}
and the subscript $``0"$ denotes expectation values with respect to $\tilde{\cH}_0$. The self-energy is given by
\begin{align*}
\Sigma(\sigma_1,\sigma_2)&=\sum_{k\alpha}V_{k\alpha}(\sigma_1)V^*_{k\alpha}(\sigma_2)G_{k\alpha}(\sigma_1,\sigma_2)\\
&\equiv\sum_{\alpha}\Sigma_{\alpha}(\sigma_1,\sigma_2),\numb
\end{align*}
in terms of the bare lead Green's function
\begin{align}
G_{k\alpha}(\sigma_1,\sigma_2)=-\text{i}\left\langle T_{C}c_{k\alpha}(\sigma_1)c^\dag_{k\alpha}(\sigma_2) \right\rangle_0.
\end{align}
Having obtained a Dyson equation, the current expectation value can now be calculated in a straightforward manner.\cite{jau94} Specifically, by comparing the perturbation expansions, the mixed Green's function is found to be related to the dot and lead Green's functions by
\begin{align}
F_{k\alpha}(\tau,\tau')=\int_{C}\text{d}\sigma V^*_{k\alpha}(\sigma)G_{k\alpha}(\tau,\sigma)D(\sigma,\tau').
\end{align}
Returning to real time, this implies that the current from Eq.~\eqref{current_ev} can be written in terms of dot Green's function and self-energy,
\begin{align*}
\label{current_time}
I_{\alpha}(t)&=e\sum_{k}V_{k\alpha}(t)F^{-+}(t,t)+\text{c.c.}\\
&=2e\Re{\sum_{k}\left[\int_{-\infty}^{\infty}\text{d}sV_{k\alpha}(t)V^*_{k\alpha}(s)\check{G}_{k\alpha}(t,s)\check{D}(s,t)\right]^{-+}}\\
&=2e\Re{\int_{-\infty}^{\infty}\text{d}s\left[\Sigma_{\alpha}^{\text{R}}(t,s)D^{-+}(s,t)+\Sigma_{\alpha}^{-+}(t,s)D^{\text{A}}(s,t)\right]},
\numb
\end{align*}
In the second line, we defined the Keldysh-rotated matrix for the dot Green's function
\begin{align}\check{D}(t,t')=\begin{pmatrix}D^{\text{R}}(t,t')&D^{\text{K}}(t,t')\\0&D^\text{A}(t,t')\end{pmatrix},
\end{align}
which contains the retarded, kinetic and advanced components. The matrix $\check{G}_{k\alpha}(t,t')$ is defined analogously. In the last line we used the Langreth rule for the lesser component of a convolution.

\subsection{Floquet expansion}
As a result of the time convolution, Dyson equation \eqref{dyson_time} does not admit an analytical solution for arbitrary time dependent drives. For periodic drives, however, we can perform an expansion into Floquet modes to simplify the problem. Before we introduce a specific drive protocol, we outline the formalism here in general terms.

The effect of a periodic drive with frequency \(\Omega_\text{dr}\) on a quantum system can be pictured as adding to the system a number of ``photons'' each carrying the energy \(\Omega_\text{dr}\). In terms of a wave function \(\phi\) evolving according to the Schr\"odinger equation \(H(t)\phi(t)=\text{i}\partial_t\phi(t)\), this means that for a Hamiltonian \(H\) with time periodicity \(H(t)=H(t+\mathcal{T})\) for all \(t\),  \(\mathcal{T}=2\pi/\Omega_\text{dr}\), there is a complete set of solutions \(\phi_{\alpha}(t)\) with
\begin{align}
\phi_{\alpha}(t)=\text{e}^{\text{i}E_{\alpha}t}u_{\alpha}(t),
\end{align}
where \(u_{\alpha}\) inherits the periodicity of the Hamiltonian, \({u_{\alpha}(t+\mathcal{T})=u_{\alpha}(t)=\sum_{n\in\mathbb{Z}}\text{e}^{-\text{i}n\Omega_\text{dr}t}u_{n\alpha}}\). The Fourier-transformed Schr\"odinger equation then reads in terms of the modes \(u_{n\alpha}\),
\begin{align}
\sum_{n\in\mathbb{Z}}H_{mn}u_{n\alpha}=(E_\alpha+m\Omega_\text{dr})u_{m\alpha},
\end{align}
meaning that the time-dependent problem can be mapped onto a time-independent one involving an infinite-matrix Hamiltonian with entries \(H_{mn}=\frac{1}{\mathcal{T}}\int_{-\mathcal{T}/2}^{\mathcal{T}/2}\text{d}t\text{e}^{\text{i}(m-n)\Omega_\text{dr}t}H(t)\). The Green's function \(G(t,t')\) corresponding to the Schr\"odinger equation can also be written in this representation:\cite{tsu08} Since \(G\) depends on two times, we first define its Wigner transform by writing \(G\) as a function of the relative and average times \(t_\text{rel}=t-t'\) and \(t_\text{av}=(t+t')/2\), respectively, and then taking the Fourier transform,
\begin{align}
G(t_\text{av},\omega)=\int_{-\infty}^{\infty}\text{d}t_\text{rel}\text{e}^{\text{i}\omega t_\text{rel}}G(t_\text{av},t_\text{rel}).
\end{align}
As a consequence of the periodicity of the Hamiltonian, the Green's function is itself periodic in the average time and can hence be expanded into Fourier modes,
\begin{align}
G(n,\omega)=\frac{1}{\mathcal{T}}\int_{-\mathcal{T}/2}^{\mathcal{T}/2}\text{d}t_\text{av}\text{e}^{\text{i}n\Omega_\text{dr}t_\text{av}}G(t_\text{av},\omega).
\end{align}
Finally, the Floquet matrix Green's function is defined by
\begin{align}
G_{mn}(\omega)=G\left(m-n,\omega+\frac{m+n}{2}\Omega_\text{dr}\right).
\label{wignerfloquet}
\end{align}
This representation allows to write convolutions in time domain as matrix multiplications in frequency space: For a function \(C(t,t')=\int_{-\infty}^{\infty}\text{d}sA(t,s)B(s,t')\), the Floquet expansion is given by \(C_{mn}(\omega)=\sum_{k=-\infty}^{\infty}A_{mk}(\omega)B_{kn}(\omega)\). In order to perform this multiplication explicitly, it is useful for make an approximation by only taking a finite number \(N_\text{Fl}\) of Floquet modes into account. This corresponds to limiting the dimension of the matrices to \(N\equiv2N_\text{Fl}+1\), with the index \(k\) running from \(-N_\text{Fl}\) to \(N_\text{Fl}\).

In this manner, Eq.~\eqref{current_time} yields the Floquet components of the current expectation value,
\begin{align*}
\label{floquetcurrent}
&\left\langle I_{\alpha}(\omega)\right\rangle_{mn}\\
&=e\sum_{k=-\infty}^{\infty}\left[\left(\Sigma_{\alpha}^{\text{R}}\right)_{mk}(\omega)D^{-+}_{kn}(\omega)+\left(\Sigma_{\alpha}^{-+}\right)_{mk}(\omega)D^{\text{A}}_{kn}(\omega)+\right.\\
&+\left. \left(\Sigma_{\alpha}^{\text{R}}\right)_{-m-k}(-\omega)^*D^{-+}_{-k-n}(-\omega)^*+\left(\Sigma_{\alpha}^{-+}\right)_{-m-k}(-\omega)^*D^{\text{A}}_{-k-n}(-\omega)^* \right].\numb
\end{align*}
In particular, the total DC current through the quantum dot takes shape as
\begin{align}
\label{dc_current}
\left\langle I \right\rangle^\text{DC}=\frac{\Omega_\text{dr}}{2\pi}\int_{-\mathcal{T}/2}^{\mathcal{T}/2}\text{d}t_\text{av}\left\langle I(t_\text{av})\right\rangle=\int\frac{\text{d}\omega}{2\pi}\left\langle I(\omega)\right\rangle_{00},
\end{align}
where we defined the current difference \(I=I_\text{L}-I_{\text{R}}\). It remains to calculate the Floquet expansions of the dot Green's function and of the self-energy, which is readily achieved: Since the Dyson equation, Eq.~\eqref{dyson_time}, only involves a time convolution, its retarded and advanced components can also be written in terms of a Floquet matrix multiplication,
\begin{align*}
\label{floquetdyson}
\left(D^{\text{R,A}}\right)_{mn}(\omega)=&\left(D^{\text{R,A}}_0\right)_{mn}(\omega)\\
&+\sum_{k,l=-\infty}^{\infty}\left(D^{\text{R,A}}_0\right)_{mk}(\omega)\Sigma_{kl}^{\text{R,A}}(\omega)D^{\text{R,A}}_{ln}(\omega).\numb
\end{align*}
This equation can be explicitly solved for \(D^{\text{R,A}}\) by inversion, provided we truncate the matrices to a finite dimension \(N\). Once \(D^{\text{R,A}}\) are known, the lesser component can also be calculated using the Keldysh integral equation,
\begin{align}
D^{-+}(t,t')=\int\text{d}s\text{d}s'D^\text{R}(t,s)\Sigma^{-+}(s,s')D^\text{A}(s',t'),
\end{align}
whose Floquet expansion is again given by a matrix multiplication.

Finally, we give the components of the Dyson equation for a general form of the driving protocol used in the following sections,
\begin{align*}
\epsilon(t)&=A\cos{\Omega_\epsilon t},\\
V_{k\alpha}(t)&=v_{k\alpha}\left[1+\Delta\cos{(\Omega_V t+\phi_\alpha)}\right],\numb
\end{align*}
where the formalism admits any kind of commensurate choice for the drive frequencies \(\Omega_\epsilon\) and \(\Omega_V\). Expanding the bare Green's function, we obtain
\begin{align*}
\label{baredot}
D_0(t_\text{av},t_\text{rel})=&\sum_{n\in\mathbb{Z}}\text{e}^{\text{i}n\Omega_{\epsilon}t_\text{av}}\int_{-\infty}^{\infty}\frac{\text{d}\omega}{2\pi}\text{e}^{-\text{i}\omega t_\text{rel}}\\
&\times\sum_{m\geq0}\sum_{k=0}^{2m+n}D^\text{free}_0\left(\omega-\left(m+\frac{n}{2}-k\right)\Omega_{\epsilon}\right)\lambda_n^{mk},\numb
\end{align*}
where we abbreviated \(\lambda_n^{mk}=\frac{(-1)^k}{m!(m+n)!}\left(\frac{A}{2\Omega_{\epsilon}}\right)^{2m+n}\begin{pmatrix}2m+n\\k\end{pmatrix}\). The bare dot Green's function in the absence of drive is denoted by \(D^\text{free}_0(\omega)\), with its retarded component given by
\begin{align}
\label{freedot}
\left(D_0^\text{free}\right)^\text{R}(\omega)=\text{e}^{-\lambda^2/\Omega^2}\sum_{k\geq0}\frac{\lambda^{2k}/\Omega^{2k}}{k!}\frac{1}{\omega-\epsilon-k\Omega+\text{i}0^+}.
\end{align}
Hence, the bare dot Green's function is given by a series of resonances at integer multiples of the vibron frequency, which follow a drive-modified Poisson distribution strongly dependent on the electron-vibron coupling parameter \(\lambda/\Omega\).

Taking the wide-band limit for the leads and their coupling to the dot results in a frequency-independent bare electronic tunneling rate \(\Gamma=2\pi\sum_{k}|v_{k\alpha}|^2\delta(\omega-\epsilon_{k\alpha})\), and yields the mode expansion for the retarded self-energy,
\begin{align}
\label{sigma_r_wigner}
\Sigma^{\text{R}}_{\alpha}(n,\omega)=-\text{i}\frac{\Gamma_\alpha}{2}\begin{cases}\frac{|\Delta|^2}{4}\text{e}^{-\text{i}2\phi_\alpha},&n=-2\\
\frac{\Delta+\Delta^*}{2}\text{e}^{-\text{i}\phi_\alpha},&n=-1\\
1+\frac{|\Delta|^2}{2},&n=0\\
\frac{\Delta+\Delta^*}{2}\text{e}^{\text{i}\phi_\alpha},&n=1\\
\frac{|\Delta|^2}{4}\text{e}^{\text{i}2\phi_\alpha},&n=2
\end{cases},
\end{align}
where the limitation of the Wigner mode indices \(n\in\left\{0,\pm1,\pm2\right\}\) results from the single-mode character of \(V_{k\alpha}\). Furthermore, the lesser component reads
\begin{align*}
\label{sigma_lesser_wigner}
&\Sigma^{-+}_{\alpha}(n,\omega)=\\
=&\text{i}\Gamma_\alpha\begin{cases}\frac{|\Delta|^2}{4}\text{e}^{-\text{i}2\phi_\alpha}n_{\text{F}\alpha}(\omega),&n=-2\\
\frac{\Delta}{2}\text{e}^{-\text{i}\phi_\alpha}n_{\text{F}\alpha}(\omega-\Omega_V/2)+\frac{\Delta^*}{2}\text{e}^{-\text{i}\phi_\alpha}n_{\text{F}\alpha}(\omega+\Omega_V/2),&n=-1\\
n_{\text{F}\alpha}(\omega)+\frac{|\Delta|^2}{4}n_{\text{F}\alpha}(\omega+\Omega_V)+\frac{|\Delta|^2}{4}n_{\text{F}\alpha}(\omega-\Omega_V),&n=0\\
\frac{\Delta^*}{2}\text{e}^{\text{i}\phi_\alpha}n_{\text{F}\alpha}(\omega-\Omega_V/2)+\frac{\Delta}{2}\text{e}^{\text{i}\phi_\alpha}n_{\text{F}\alpha}(\omega+\Omega_V/2),&n=1\\
\frac{|\Delta|^2}{4}\text{e}^{\text{i}2\phi_\alpha}n_{\text{F}\alpha}(\omega).&n=2
\end{cases}\numb
\end{align*}
Here, we consider the leads at zero temperature, meaning that the Fermi function of lead \(\alpha\) is a step function \({n_{\text{F}\alpha}(\omega)=\theta(\mu_\alpha-\omega)}\), with the chemical potentials \(\mu_\alpha\).

After converting these Wigner expansions into Floquet modes via Eq.~\eqref{wignerfloquet}, the Dyson equation Eq.~\eqref{floquetdyson} can be solved to obtain the renormalized dot Green's function and hence the current. A detailed derivation of Eqs.~\eqref{baredot}--\eqref{sigma_lesser_wigner} can be found in Appendix \ref{dyson_app}.

\section{Current under bias}
\label{sec_bias}
In this section, we use the Floquet-Green's function expression for the calculation of the average current. Specifically, we apply it to a polaron quantum dot subject to a bias \(eV\equiv\mu_\text{L}-\mu_\text{R}\) between left and right leads, with a time-dependent drive applied to both the dot energy and the coupling to the leads. By comparing to previous results,\cite{hau16} we are also able to estimate the influence of Floquet harmonics beyond the leading order on the lifting of Franck-Condon blockade.

\begin{figure}[t]
    \includegraphics[width=.99\columnwidth]{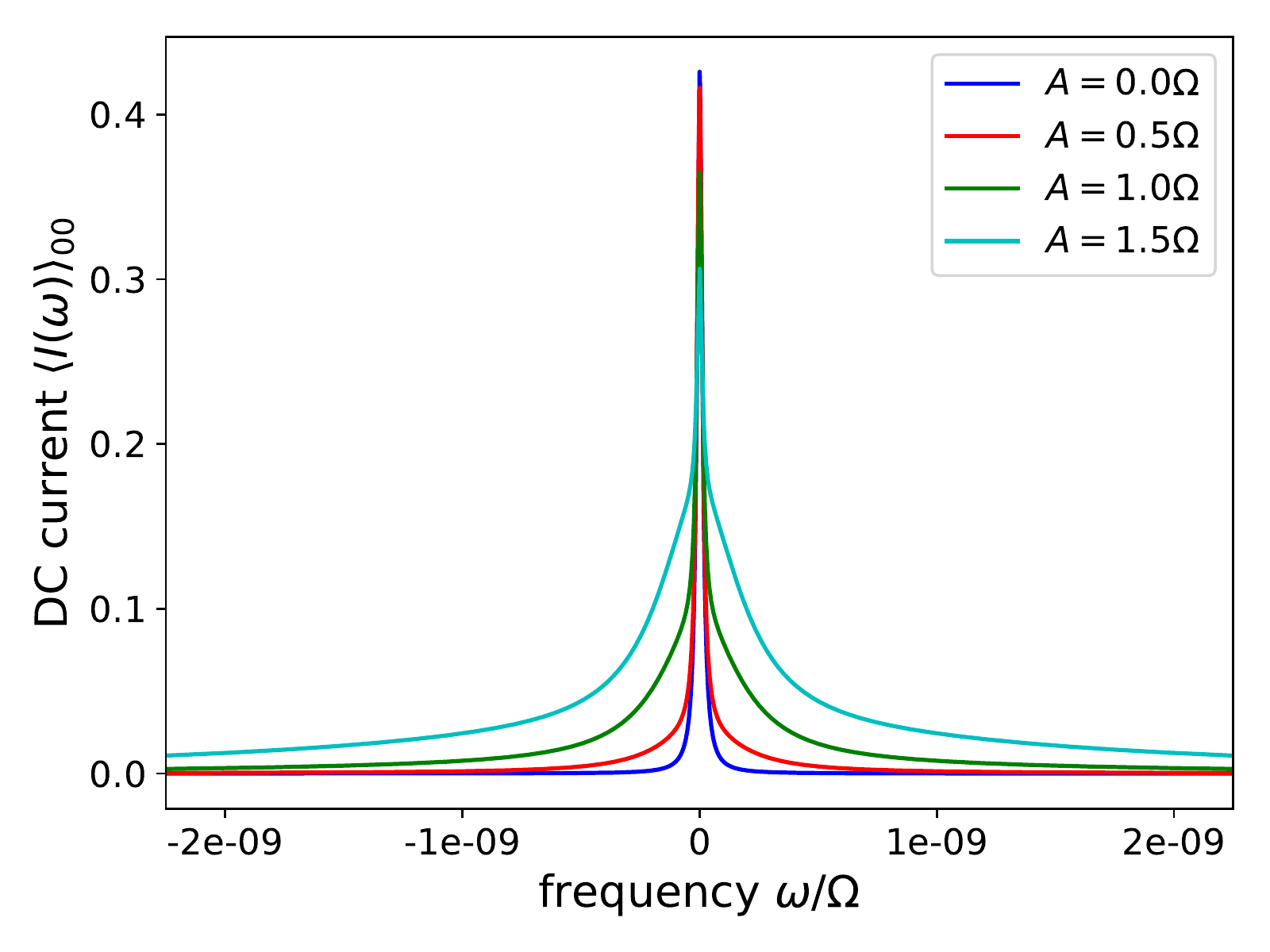}
    \caption{Frequency-resolved DC current integrand \(\left\langle I_{\alpha}(\omega)\right\rangle_{00}\) through a biased polaron dot (\(\mu_\text{L}=-\mu_\text{R}=-\Omega/4\)), evaluated near the zeroth-order resonance. We use Eq.~\ref{floquetcurrent} with Floquet matrix dimension \(N_\text{Fl}=5\) for different values of the dot drive amplitude \(A\), and coupling drive amplitude \(\Delta=0.6\). Increasing \(A\) causes substantial widening of the peak, signifying strong current increase.}
    \label{biasintegrand_nophase}
\end{figure}

\begin{figure}[t]
    \includegraphics[width=.99\columnwidth]{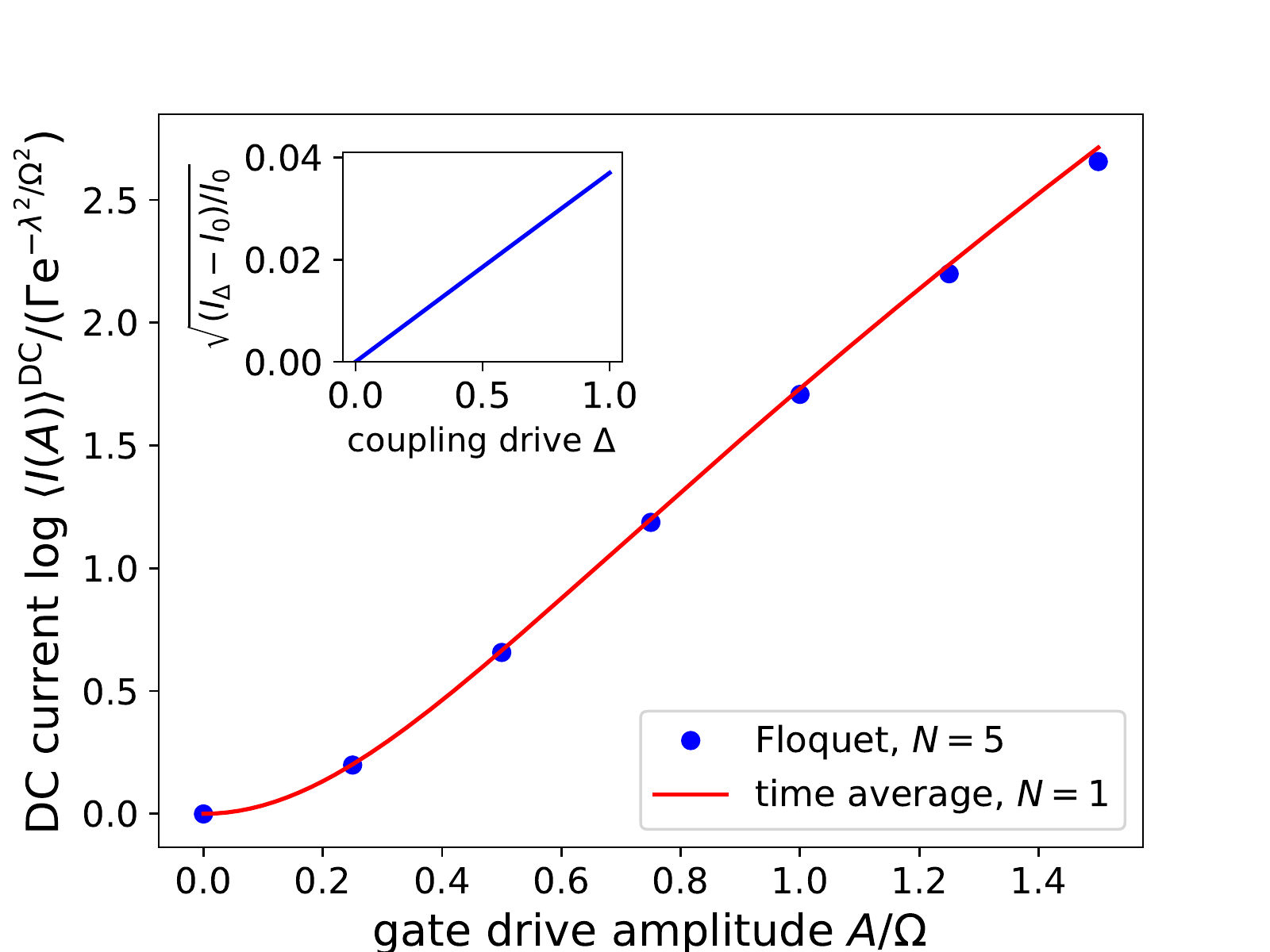}
    \caption{Exponential FCB (\(\lambda=4\Omega\)) lifting by driving the dot energy in the presence of a bias window \(\mu_\text{L}=-\mu_\text{R}=\Omega/4\). Floquet result (blue) with dimension \(N_\text{Fl}=5\) to be compared with the result (red) from Ref.~[\onlinecite{hau16}], corresponding to \(N_\text{Fl}=1\). \emph{Inset}: Increasing the coupling drive for fixed dot drive leads to a small-scale quadratic rise in tunneling current \(I_\Delta\).}
    \label{zerocurrent_dot_coupling}
\end{figure}

The pumping protocol examined here consists of two single-mode drives,
\begin{align*}
\epsilon(t)&=A\cos{\Omega t},\\
V_{k\alpha}(t)&=v_{k\alpha}\left(1+\Delta\cos{\Omega t/2}\right).\numb
\end{align*}
The dot drive \(\epsilon(t)\) is chosen to be resonant with the vibron mode so as to maximize the resulting current amplification.\cite{hau16} Moreover, the coupling is driven at half of this frequency, so that the self-energy, which contains the square of the coupling, is itself in resonance with the dot and the vibron. According to Eq.~\eqref{freedot}, the bare dot Green's function features resonances at all positive integer multiples of \(\Omega\). We choose a bias voltage $V$ in such a way as to reach the regime which exhibits the strongest Franck-Condon blockade as well as the most pronounced current response to drive,\cite{hau16} i.e., \(\Gamma\ll eV\ll\Omega\).
The integrand \(\left\langle I_{00}(\omega)\right\rangle\) that gives rise to the DC current flowing through this setup can be seen in Fig.~\ref{biasintegrand_nophase} for different values of the dot drive amplitude \(A\). All current is due to the single resonance within the bias window, whose width strongly increases with \(A\). The drive dependence of the integrated DC current \(\langle I\rangle^\text{DC}\) calculated from Eq.~\eqref{dc_current} is illustrated in Fig.~\ref{zerocurrent_dot_coupling}

The striking feature here is the degree of difference between the respective current responses to the dot and coupling drives: The dot drive causes exponential lifting of the Franck-Condon blockade as a function of dot drive amplitude \(A\), regardless of the presence of the coupling drive. By contrast, the dependence on the coupling drive amplitude \(\Delta\) is only quadratic, as is expected from the fact that it directly multiplies the tunneling coefficient \(v_{k\alpha}\), the square of which is featured in the bare electronic tunneling rate \(\Gamma\). Compared to the dependence on the dot drive, the effect of the coupling drive is minuscule: Increasing the amplitude up to the static  value \(v_{k\alpha}\) of the coupling causes a mere \(2\%\) difference in current.

Finally, Fig.~\ref{zerocurrent_dot_coupling} also shows that the impact of higher Floquet harmonics on the current response is negligible: The Floquet result obtained using a matrix dimension \(N_\text{Fl}=5\) in Eq.~(\ref{floquetcurrent}) differs little from the outcome of the simplified calculation from Ref.~[\onlinecite{hau16}], where the response to the dot drive was calculated using only time-averaged Green's functions, which is equivalent to truncating the Floquet matrices down to \(N_\text{Fl}=1\).

\section{Polaron pumping}
\label{sec_pumping}
\begin{figure}[t]
    \includegraphics[width=.99\columnwidth]{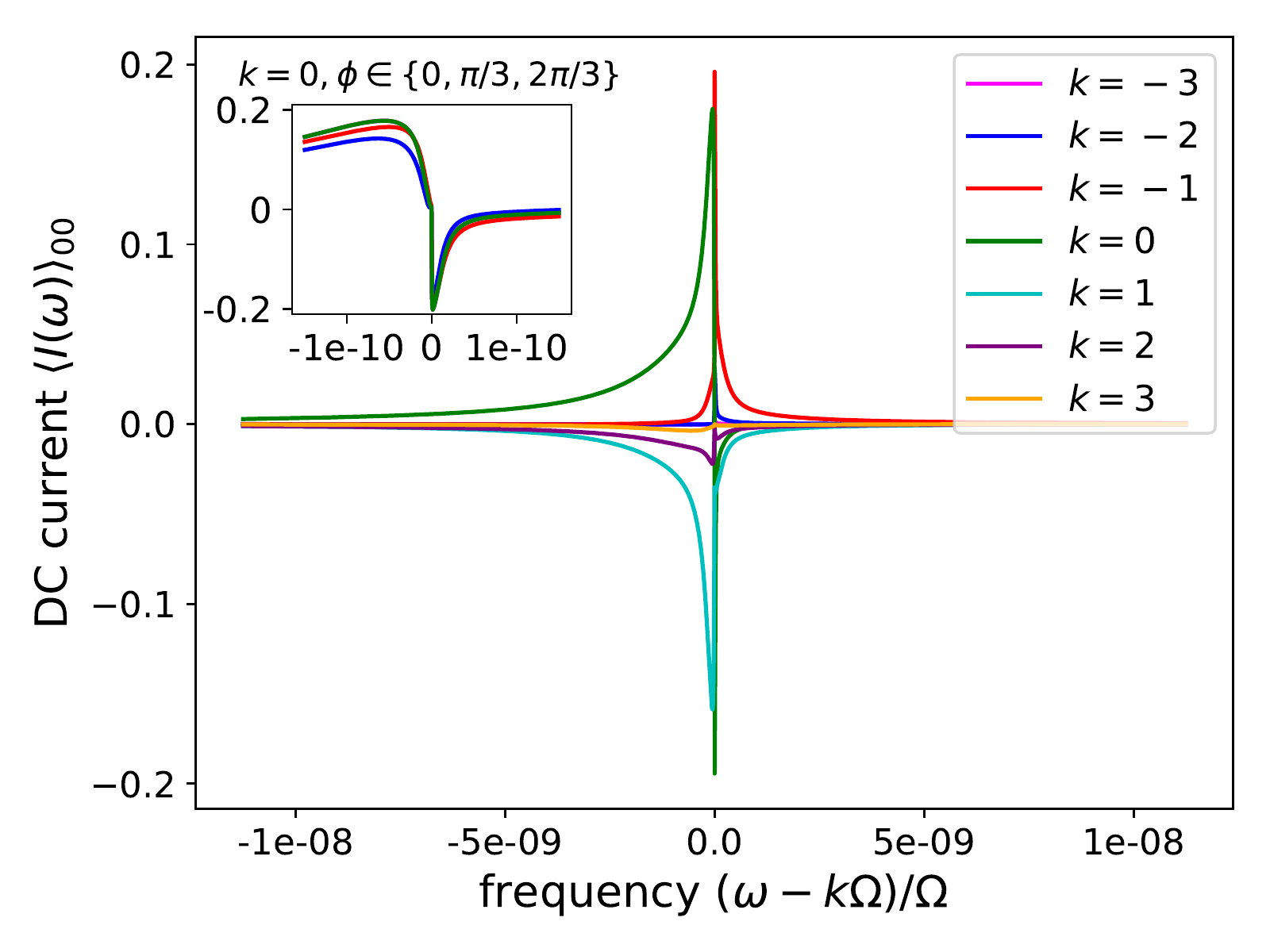}
    \caption{Frequency-resolved DC current integrand \(\left\langle I_{\alpha}(\omega)\right\rangle_{00}\), through an unbiased polaron dot, evaluated near the peak locations (peak order \(k\)), using Eq.~(\ref{floquetcurrent}) with Floquet matrix dimension \(N_\text{Fl}=21\). The side resonances appearing outside the bias window are comparable in size to the central one at \(k=0\), but different sides lead to partial cancellation upon integration. \emph{Inset}: Modification of the shape of the central resonance for different values of phase shift.}
    \label{nobiasintegrand}
\end{figure}

In this section we consider the unbiased polaron dot, i.e., we set the chemical potentials of the leads to \(\mu_\text{L}=\mu_\text{R}=0\). In this case, a DC current can still flow in the presence of a drive protocol which breaks the left-right symmetry. In order to break this symmetry, we add phase differences \(\phi_\alpha\) to the left and right coupling drives, so that the driving protocol is given by
\begin{align*}
\epsilon(t)&=A\cos{\Omega t},\\
V_{k\alpha}(t)&=v_{k\alpha}\left[1+\Delta\cos{(\Omega t/2+\phi_\alpha)}\right].\numb
\end{align*}
For such a setup, several channels contribute to charge transport through the dot, as evidenced by the current integrand depicted in Fig.~\ref{nobiasintegrand}, which exhibits several resonances of comparable height. The role of higher Floquet harmonics is evident from the slower convergence of the result as a function of the truncation index \(N_\text{Fl}\).

\begin{figure}[t]
     \includegraphics[width=0.99\columnwidth]{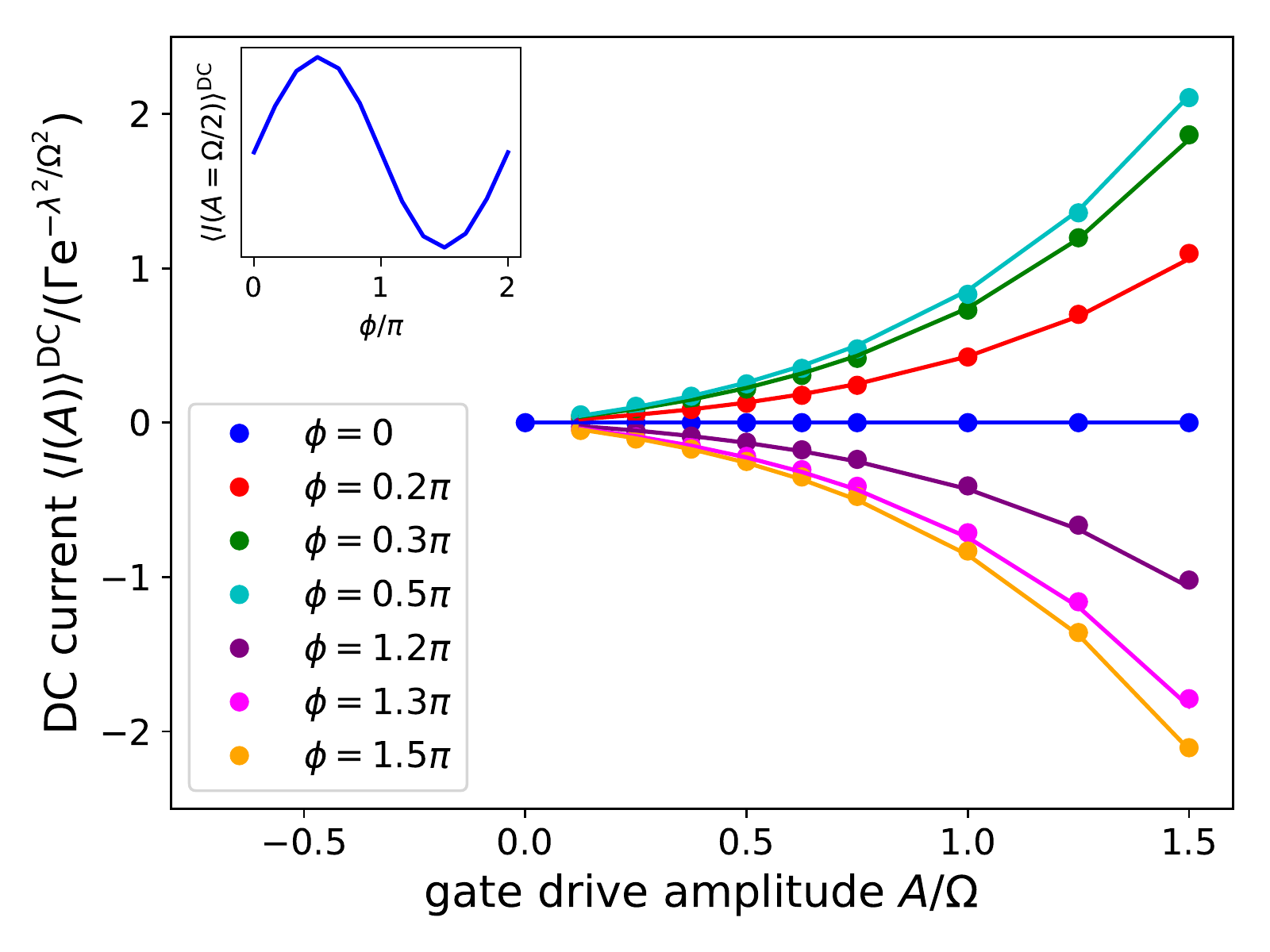}
  \caption{DC current \(\left\langle I \right\rangle^\text{DC}\) as a function of the dot drive amplitude \(A\) for different values of phase difference \(\phi_\text{L}=-\phi_\text{R}=\phi/2\), and fixed coupling drive amplitude \(\Delta=0.6\). \emph{Dots}: simulation of Eq.~\eqref{dc_current}, \emph{lines}: numerical fit proportional to \(\frac{A}{\Omega}\text{e}^{A/\Omega}\sin{\phi}\). The low-amplitude regime behaves in a way similar to purely electronic pumping, with exponential lifting of FCB (\(\lambda=4\Omega\)) evident for larger amplitude. \emph{Inset}: Sinusoidal phase dependence of \(\left\langle I \right\rangle^\text{DC}\) for fixed value of \(A\).}
\label{zerocurrent}
\end{figure}

The interplay between the drive parameters is more complex than in the biased case: Fundamentally, there can be no current without breaking of left-right symmetry, so the phase differences are essential to achieve current flow. In the same vein, the dot drive \(\epsilon(t)\) by itself will not produce any current: the coupling drive \(V_{k\alpha}(t)\) is also required. This is reflected in Fig.~\ref{zerocurrent}, where dependencies of the DC current \(\langle I\rangle^\text{DC}\) through the unbiased dot on dot drive and phase shift are on display, with the dependence on the phase difference isolated in the inset: In the absence of dot drive (\(A=0\)), no current is measured, independently of coupling drive and phase, and the same is true for the case of zero phase difference (\(\phi_\text{L}=\phi_\text{R}\)). Even though the phase difference appears to have little influence on the shape of the resonances, it has a much more significant effect on the integrated current: For a nonvanishing phase difference, a current response to the dot drive is observed, with the current increasing roughly linearly in the regime of \(A\ll\Omega\), and in a superlinear fashion for larger values of \(A\). On the other hand, for a fixed value of \(A\), the current depends on the phase difference in a sinusoidal fashion.  Taken together, we find that the current response to the driving protocol is given by
\begin{align}\label{eq:I_drive}
\left\langle I(A,\Delta) \right\rangle^\text{DC}\propto\Gamma\text{e}^{-\lambda^2/\Omega^2}\frac{A}{\Omega}\text{e}^{A/\Omega}\Delta^2\sin{\phi},
\end{align}
where $\phi = \phi_\text{L}-\phi_\text{R}$, for small values of \(\Delta\).

Equation (\ref{eq:I_drive}) is the main result of this work. Both the linear factor of \(A\) and the sine of the phase difference are known features of pumping in non-interacting, purely electronic systems.\cite{arr08} The dependence on \(\sin\phi\) appears to be a feature of adiabatic pumping\cite{wan02} persisting in the highly non-adiabatic protocol considered here. On the other hand,  the exponential factor is familiar from the lifting of Franck-Condon blockade in the unbiased case of our model.\cite{hau16} Hence we conclude that the polaron dot exhibits pumping characteristics similar to a non-interacting system, in addition to Franck-Condon blockade and strong amplification of pump current by application of an AC gate voltage.

\section{Conclusion}
\label{sec_conclusion}
We have examined the interplay between strong electromechanical coupling and periodic drive protocols in a model of a quantum dot coupled to a pair of metallic leads, subject to AC gate voltages modulating both the dot energy and the coupling to the leads. By combining a perturbative approach in the tunnel coupling with an expansion in Floquet modes, we were able to quantify the effects of multi-parameter drives and of higher drive harmonics on the DC current through the quantum dot.

Our main findings are twofold: Firstly, we studied the case of a biased quantum dot in the limit of bias voltage small compared to the vibron frequency. In this situation, the dominant effect is the lifting of the polaron-induced Franck-Condon blockade as a result of driving the dot energy in resonance with the vibrational mode on the dot. In particular, the coupling drive only has a minimal effect on the current in this regime, and the current response to the drive is well approximated by the Floquet component which encodes time-averaged contributions. This result also serves to confirm earlier work on drive-induced lifting of Franck-Condon blockade, where only time-averaged correlation functions were considered in the perturbative expansion.\cite{hau16}

Secondly, we applied the same formalism to the unbiased quantum dot. There, the interplay between both dot and coupling drives makes it possible to pump a DC current through the system: Similar to the case of charge pumping in purely electronic systems, we find that in the regime of weak dot drive the  DC current flow is approximately proportional to the drive amplitude, as long as a phase difference between the left and right coupling drives is employed to break left-right symmetry. The dependence of the current on this phase is found to always be sinusoidal, irrespective of the dot drive amplitude. In the regime of strong dot drive, in contrast, the current response becomes superlinear and approximates the exponential characteristic found in the biased quantum dot. Thus, the driven unbiased polaron quantum dot combines the exponential lifting of Franck-Condon blockade with features of adiabatic charge pumping through purely electronic systems, even far away from the adiabatic limit.

As recent experiments have used electronic gates to localize quantum dots on carbon nanotubes, we anticipate that these could also be employed to supply the AC voltages we use to predict charge pumping, thus providing an implementation of our model and adding to the versatility of carbon nanotubes as elements of circuitry. Moreover, the Anderson-Holstein Hamiltonian used in this work is a fairly general model and could be realized in multiple ways, as long as there is a way to engineer strong coupling between a fermion and a bosonic mode and subject this system to several resonant drives. In particular, an optomechanical implementation could be envisioned, with cavity modes taking the role of the drive.

The model itself can be extended to include multiple electron levels and oscillator modes by replacing the expression for the dot Green's function by a more complex one; this appears as a promising way to better capture the possible complexities of experiments. Furthermore, the generality of the Floquet formalism also permits the consideration of more complex driving schemes as well as the time-dependent current response. Finally, while the polaron tunneling approximation allows to simplify the diagrammatic expansion substantially and yet capture the effect of electron-phonon interaction in the parameter ranges considered in this work, it would be interesting to explore alternatives such as the Floquet DMFT,\cite{tsu08} which would allow to move beyond this approximation, as well as to consider more complex system Hamiltonians.

\acknowledgments
P.H. and T.L.S. would like to acknowledge financial support by the National Research Fund Luxembourg (ATTRACT 7556175) and discussions with M.~Esposito. J.G. is supported by Singapore MOE Academic Research Fund Tier-2 project (Project No. MOE2014-T2-2- 119 and WBS No. R-144-000-350-112).

\appendix
\section{Bare Green's functions}
\label{dyson_app}
Here, we derive the expressions Eq.~\eqref{baredot} and Eq.~\eqref{freedot} for the bare and free dot Green's functions. The free dot Green's function is calculated by evolution with the time independent Hamiltonian \(\cH_0^\text{free}=\epsilon d^\dag d +\Omega a^\dag a\) and factors into electronic and vibrational parts,
\begin{align}
\label{freegf}
D_0^\text{free}(\tau,\tau')=-\text{i}\left\langle T_{C}d(\tau)d^\dag(\tau')\right\rangle_0^\text{free}\left\langle T_{C}X(\tau)X^\dag(\tau')\right\rangle_0^\text{free},
\end{align}
where \(X(\tau)=\text{e}^{-\frac{\lambda^2}{\Omega^2}\left(a^\dag\text{e}^{\text{i}\Omega \tau}-a\text{e}^{-\text{i}\Omega \tau}\right)}\). Using Keldysh matrix notation, the electronic Green's function takes shape as
\begin{align}
& \left\langle T_{C}d(\tau)d^\dag(\tau')\right\rangle_0^\text{free} =\notag \\
& \mathrm{e}^{-\mathrm{i}\epsilon(t-t^{\prime})}\begin{pmatrix}-n_\text{d}+\theta(t-t^{\prime})&&-n_\text{d}\\1-n_\text{d}&&-n_\text{d}+\theta(t^{\prime}-t)\end{pmatrix}.
\end{align}
Throughout this work, we focus on times beyond the transient regime, meaning the effect of the initial dot occupation \(n_\text{d}\) is negligible and it can hence be set to \(n_\text{d}=0\). On the other hand, the vibrational part reads
\begin{align}
\left\langle T_{C}X(\tau)X^\dag(\tau')\right\rangle_0^\text{free}=\text{e}^{-\lambda^2/\Omega^2}\begin{pmatrix}\text{e}^{\frac{\lambda^2}{\Omega^2}\text{e}^{-\text{i}\Omega|t|}}&\text{e}^{\frac{\lambda^2}{\Omega^2}\text{e}^{\text{i}\Omega t}}\\
\text{e}^{\frac{\lambda^2}{\Omega^2}\text{e}^{-\text{i}\Omega t}}&\text{e}^{\frac{\lambda^2}{\Omega^2}\text{e}^{\text{i}\Omega|t|}}\end{pmatrix}.
\end{align}
The retarded component in Eq.~\eqref{freedot} is obtained by Fourier transform of Eq.~\eqref{freegf} and using the definition \({\left(D_0^\text{free}\right)^\text{R}_0=\left(D_0^\text{free}\right)^{--}-\left(D_0^\text{free}\right)^{-+}}\).

The bare dot Green's function differs from the free one by the additional evolution with the drive Hamiltonian \(\cH_\text{dr}(t)=A\cos{\Omega_\epsilon t}\), which implies
\begin{align*}
D_0(\tau,\tau')&=D_{0}^\text{free}(\tau,\tau')\text{e}^{-\text{i}\frac{A}{\Omega_{\epsilon}}\left(\sin{\Omega_{\epsilon}\tau}-\sin{\Omega_{\epsilon}\tau'}\right)}\\
&=\sum_{n\in\mathbb{Z}}\text{e}^{\text{i}n\Omega_{\epsilon}\tau_\text{ave}}D^\text{free}_0(\tau_\text{rel})\text{i}^nJ_n\left(-\frac{2A}{\Omega_{\epsilon}}\sin{\frac{\Omega_\epsilon \tau_\text{rel}}{2}}\right),\numb
\end{align*}
where \(J_n\) denotes the \(n\)-th Bessel function of the first kind.
By Fourier transforming in the relative time coordinate, we obtain the Wigner expansion,
\begin{widetext}
\begin{align*}
&\int_{-\infty}^\infty\text{d}\tau_\text{rel}\text{e}^{\text{i}\omega\tau_\text{rel}}\text{i}^nJ_n\left(-\frac{2A}{\Omega_{\epsilon}}\sin{\frac{\Omega_{\epsilon}\tau_\text{rel}}{2}}\right)=\int_{-\infty}^\infty\text{d}\tau_\text{rel}\text{e}^{\text{i}\omega\tau_\text{rel}}\text{i}^n\sum_{m\geq0}\frac{(-1)^m}{m!(m+n)!}\left(-\frac{A}{\Omega_{\epsilon}}\sin{\frac{\Omega_{\epsilon}\tau_\text{rel}}{2}}\right)^{2m+n}\\
&=\int_{-\infty}^\infty\text{d}\tau_\text{rel}\text{e}^{\text{i}\omega\tau_\text{rel}}\text{i}^n\sum_{m\geq0}\frac{(-1)^m}{m!(m+n)!}\left(-\frac{A}{\text{i}2\Omega_{\epsilon}}\right)^{2m+n}\left(\text{e}^{\text{i}\frac{\Omega_{\epsilon}\tau_\text{rel}}{2}}-\text{e}^{-\text{i}\frac{\Omega_{\epsilon}\tau_\text{rel}}{2}}\right)^{2m+n}\\
&=\int_{-\infty}^\infty\text{d}\tau_\text{rel}\sum_{m\geq0}\sum_{k=0}^{2m+n}\text{e}^{\text{i}\left(\omega+k\Omega_{\epsilon}-m\Omega_{\epsilon}-\frac{n}{2}\Omega_{\epsilon}\right)\tau_\text{rel}}\frac{(-1)^k}{m!(m+n)!}\left(\frac{A}{2\Omega_{\epsilon}}\right)^{2m+n}\begin{pmatrix}2m+n\\k\end{pmatrix}\\
&=2\pi\sum_{m\geq0}\sum_{k=0}^{2m+n}\delta\left(\omega-\left(m+\frac{n}{2}-k\right)\Omega_{\epsilon}\right)\frac{(-1)^k}{m!(m+n)!}\left(\frac{A}{2\Omega_{\epsilon}}\right)^{2m+n}\begin{pmatrix}2m+n\\k\end{pmatrix}.\numb
\end{align*}
\end{widetext}
This expression is then convolved with \(D^\text{free}_0(\omega)\):
\begin{align}
D_0(t_\text{av},t_\text{rel})=\sum_{n\in\mathbb{Z}}\text{e}^{\text{i}n\Omega_{\epsilon}t_\text{av}}\int_{-\infty}^{\infty}\frac{\text{d}\omega}{2\pi}\text{e}^{-\text{i}\omega t_\text{rel}}\notag \\
\sum_{m\geq0}\sum_{k=0}^{2m+n}D^\text{free}_0\left(\omega-\left(m+\frac{n}{2}-k\right)\Omega_{\epsilon}\right)\lambda_n^{mk},
\end{align}
with 
\begin{align}
  \lambda_n^{mk}=\frac{(-1)^k}{m!(m+n)!}\left(\frac{A}{2\Omega_{\epsilon}}\right)^{2m+n}\begin{pmatrix}2m+n\\k\end{pmatrix}.
\end{align}

\section{Self-energy}
\label{selfenergy_app}
In the following, we give the derivations of the mode expansions  Eq.~\eqref{sigma_r_wigner} and Eq.~\eqref{sigma_lesser_wigner} for the retarded and lesser component of the lead self-energy, respectively. Resummation of the perturbation series for \(D(\tau,\tau')\) produces the Dyson equation \eqref{dyson_time}, with a self-energy \(\Sigma\) which contains the time dependence of the coupling \(V_{k\alpha}(t)=v_{k\alpha}\left[1+\Delta\cos{(\Omega_V t+\phi_\alpha)}\right]\),
\begin{align}
\Sigma(\tau_1,\tau_2)=\sum_{k\alpha}V_{k\alpha}(\tau_1)V^*_{k\alpha}(\tau_2)G_{k\alpha}(\tau_1,\tau_2).
\end{align}

Introducing Wigner coordinates and Fourier transforming, we obtain
\begin{widetext}
\begin{align*}
&\Sigma(t_\text{av},\omega)_{\alpha}=\sum_{k}|v_{k\alpha}|^2\int_{-\infty}^{\infty}\text{d}t_\text{rel}\text{e}^{\text{i}\omega t_\text{rel}}V_{k\alpha}(t_\text{av}+t_\text{rel}/2)V^*_{k\alpha}(t_\text{av}-t_\text{rel}/2)G_{k\alpha}(t_\text{rel})\\
&=\sum_{k}|v_{k\alpha}|^2\int_{-\infty}^{\infty}\text{d}t_\text{rel}\text{e}^{\text{i}\omega t_\text{rel}}\left[\frac{|\Delta|^2}{4}\text{e}^{-\text{i}2\phi_\alpha}G_{k\alpha}(t_\text{rel})\text{e}^{-\text{i}2\Omega_V t_\text{av}}+\frac{|\Delta|^2}{4}\text{e}^{\text{i}2\phi_\alpha}G_{k\alpha}(t_\text{rel})\text{e}^{\text{i}2\Omega_V t_\text{av}}+\left(1+\frac{|\Delta|^2}{4}\text{e}^{\text{i}\Omega_V t_\text{rel}}+\frac{|\Delta|^2}{4}\text{e}^{-\text{i}\Omega_V t_\text{rel}}\right)G_{k\alpha}(t_\text{rel})+\right.\\
&\qquad\qquad+\left.\left(\frac{\Delta}{2}\text{e}^{\text{i}\left(\Omega_V t_\text{rel}/2+\phi_\alpha\right)}+\frac{\Delta^*}{2}\text{e}^{\text{i}\left(-\Omega_V t_\text{rel}/2+\phi_\alpha\right)}\right)G_{k\alpha}(t_\text{rel})\text{e}^{\text{i}\Omega_V t_\text{av}}+\left(\frac{\Delta}{2}\text{e}^{-\text{i}\left(\Omega_V t_\text{rel}/2+\phi_\alpha\right)}+\frac{\Delta^*}{2}\text{e}^{-\text{i}\left(-\Omega_V t_\text{rel}/2+\phi_\alpha\right)}\right)G_{k\alpha}(t_\text{rel})\text{e}^{-\text{i}\Omega_V t_\text{av}}\right].\numb
\end{align*}
\end{widetext}
In order to obtain the retarded and lesser components of the self energy, we substitute the free lead Green's functions
\begin{align}
G_{k\alpha}^\text{R}(t_\text{rel})&=-\text{i}\theta(t_\text{rel})\text{e}^{-\text{i}\epsilon_{k\alpha}t_\text{rel}},\notag \\
G_{k\alpha}^{-+}(t_\text{rel})&=\text{i}n_\text{F}(\epsilon_{k\alpha})\text{e}^{-\text{i}\epsilon_{k\alpha}t_\text{rel}},
\end{align}
respectively. Using the identity \(\lim_{\eta\to0^+}(\omega-\epsilon_{k\alpha}+\text{i}\eta)^{-1}=\mathcal{P}(\omega-\epsilon_{k\alpha})^{-1}-\text{i}\pi\delta(\omega-\epsilon_{k\alpha})\), with the notation \(\mathcal{P}\) for the principal value, the integral is readily performed. In the wide-band limit, the bare electronic tunneling rate \(\Gamma=2\pi\sum_{k}|v_{k\alpha}|^2\delta(\omega-\epsilon_{k\alpha})\) is set to be independent of frequency, which yields Eqs.~\eqref{sigma_r_wigner} and \eqref{sigma_lesser_wigner}.

\bibliography{pump}

\end{document}